\documentstyle[proceedings]{crckapb}

\begin{opening}
\title{THE NOVA-LIKE VARIABLES}

\subtitle{ } 

\author{Vik Dhillon}
\institute{Royal Greenwich Observatory, La Palma}

\end{opening}

\runningtitle{THE NOVA-LIKE VARIABLES}

\begin{document}

\begin{abstract}
We review optical observations and theoretical models of the non-magnetic
nova-like variables (UX UMa, VY Scl and SW Sex stars). A brief discussion of
the classification scheme is followed by a statistical overview of the observed
properties. The most important features of each of the sub-classes are then
reviewed, concluding with a summary of the theoretical models invoked to
understand these systems. 
\end{abstract}

\section{Classification}

Nova-like variables are defined as cataclysmic variable stars (CVs) which have
never been observed to undergo nova or dwarf nova-type outbursts. Such a crude
definition encompasses a wide variety of objects which can be divided into two
distinct groups; those which are believed to accrete via magnetic field lines
-- AM Her stars (or polars), DQ Her stars and intermediate polars -- and those
which are believed to accrete via an accretion disc -- UX UMa stars, VY Scl
stars (or anti-dwarf novae), SW Sex stars and AM CVn stars (or
double-degenerates). The magnetic nova-likes have been reviewed elsewhere in
this volume. The subject of this review will be the optical characteristics of
non-magnetic nova-likes (NMNLs), with the exception of the AM CVn stars which
are outside the scope of this paper. 

There is little agreement on the classification scheme for NMNLs in the
literature. Perhaps the most generally accepted scheme adheres to the 
following rules: if a nova-like shows no evidence for magnetic accretion and 
is not recognized to be a double-degenerate, it is classed as a UX UMa star. 
A UX UMa star which is observed to show states of low brightness becomes a VY 
Scl star. A number of UX UMa and VY Scl stars with high inclinations and 
periods of 3--4 hr, show single-peaked emission lines which remain (largely) 
unobscured during primary eclipse and which exhibit transient absorption 
features and distorted radial velocity curves. These systems were classed as 
SW Sex stars by Thorstensen et al. (1991). This classification is 
controversial, however, since SW Sex stars can only be recognized by their 
spectroscopic properties, whereas other classes of CV are traditionally 
classified by their photometric variations. Amongst other things, this can lead
to confusion, since a star can then be both a VY Scl star and an SW Sex star. 
In addition, it is possible that the SW Sex stars are simply the 
high-inclination counterparts of other NMNLs, in which case a separate 
classification is not justified. Bearing these caveats in mind, it is 
nevertheless sometimes convenient to be able to refer to these objects as SW 
Sex stars and this term will be used in the present review.

\section{Statistical Overview}

\begin{figure}[t]
\begin{picture}(100,186)(50,85)
\put(0,0){\includegraphics{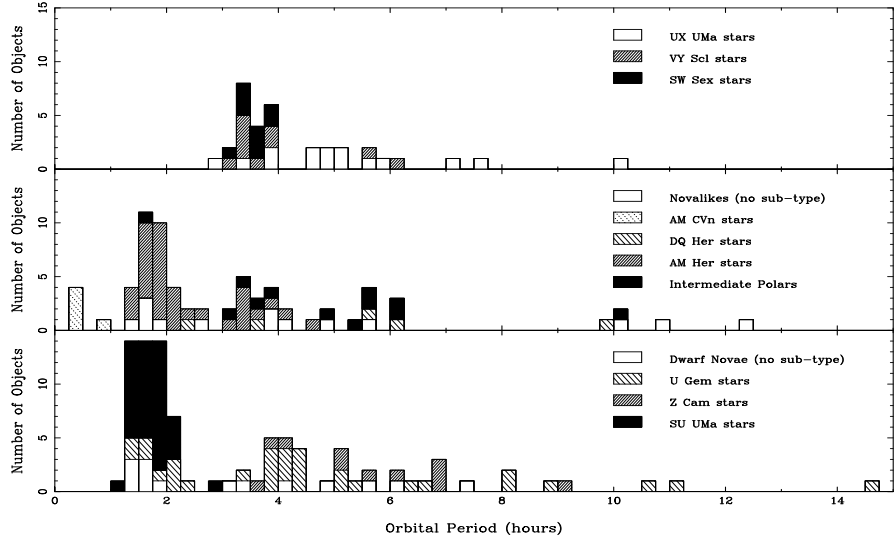}}
\noindent
\end{picture}
\caption{The orbital period distribution of CVs. Data from Ritter \&\ Kolb
(1993).} 
\end{figure}
Of the 703 CVs known, 104 are classed as nova-like variables (Downes \&\ Shara
1993). Of these, 21 are UX UMa stars and 13 VY Scl stars. Arguably, up to 6 of
these UX UMa stars and 3 of these VY Scl stars exhibit the SW Sex phenomenon
(Ritter \&\ Kolb 1993). When plotted in galactic coordinates (la Dous 1993a),
it is immediately apparent that nova-likes are found in equal numbers in the
galactic plane and at high galactic latitudes, with no tendency for their
numbers to increase towards the galactic centre. The implication is that, like
the dwarf novae, nova-likes are nearby and, given their low apparent
brightness, they must also be intrinsically faint. These conclusions are
confirmed by determinations of distances (of order hundreds of parsecs) and
absolute magnitudes ($M_V\sim4$) of nova-likes (Warner 1987). The actual
galactic distributions and space densities are, however, much more contentious
subjects -- see la Dous (1993a). 

Figure~1 shows the orbital period distribution of CVs. It can be seen that the
NMNLs are tightly grouped in the 3--4 hr period range and there are no NMNLs
below the 2--3 hr period gap. The dwarf novae, on the other hand, are found
both above and below the period gap and there appears to be a dearth of dwarf
novae in the 3--4 hr period range favoured by the NMNLs. The implications of
this result will be discussed in section~4. 

\begin{figure}[t]
\begin{picture}(100,137)(50,90)
\put(0,0){\includegraphics{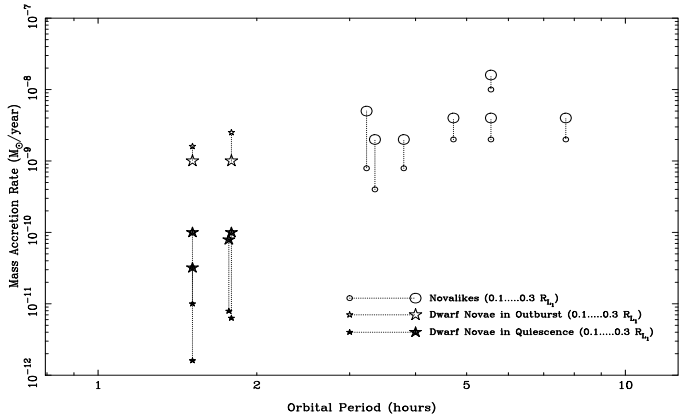}}
\noindent
\end{picture}
\caption{The dependence of mass accretion rate at disc radii 0.1 R$_{L_1}$ and
0.3 R$_{L_1}$ on orbital period for dwarf novae (in outburst and quiescence)
and NMNLs.} 
\end{figure}
One of the most important distinctions between NMNLs and other CVs can be found
by comparing their mass accretion rates. Following Horne (1993), figure~2 plots
the mass accretion rate derived from eclipse mapping experiments as a function
of binary period for a number of NMNLs and dwarf novae in outburst and
quiescence. It is evident that both NMNLs and dwarf novae in outburst accrete
mass at similar rates and at a much higher rate than dwarf novae in quiescence.
The accretion rate also appears to increase slightly with orbital period.

A discrepancy between steady-state accretion (in which the mass accretion rate
is constant throughout the disc) and the observed accretion rates can also be
inferred from figure~2. The mass accretion rate appears to decrease slightly
with radius in the dwarf novae discs during outburst. This is expected since
material should be draining from the disc as it declines from outburst. The
NMNLs, however, all show mass accretion rates that increase with radius in the
disc. The increase is low for the longer period NMNLs, but becomes significant
for systems in the 3--4 hr period range. The latter result has been inferred
from the eclipse maps of Rutten et al. (1992), which show that the radial
temperature dependences in the inner disc regions of SW Sex stars are much
flatter than the $T \propto R^{-3/4}$ relation predicted by steady-state
accretion theory. The implication is that the central regions of these discs
may be releasing some of the accretion energy in a non-radiative form, and it
seems likely that this result is linked to the SW Sex phenomenon which requires
an extra component of light that remains visible during primary eclipse (see
section~4). 

\section{Observed Properties}

The long-term light curves of NMNLs are as varied and complex as those
exhibited by novae and dwarf novae. For example, some VY Scl stars (eg. V794
Aql; Honeycutt et al. 1994a) show deep low states of $\sim 3$ magnitudes,
preceded by a number of shallower low states with saw-toothed light curves.
Other VY Scl stars (eg. DW UMa -- also an SW Sex star; Honeycutt et al. 1993),
show even deeper low states of much longer duration (typically years).
Inevitably, there are also VY Scl stars which show both types of behaviour (eg.
MV Lyr; Rosino et al. 1993). As well as low states, however, some NMNLs also 
show outbursts; Honeycutt et al. (1994b) observed a 1 magnitude outburst in 
the VY Scl star KR Aur, while Still et al. (1995a), observed the UX UMa-star 
RW Tri during a 3.5 magnitude outburst. It is not known how often such 
eruptions occur in NMNLs, and whether or not they are related to nova or dwarf
nova-type outbursts is unclear. 

Other types of long-term photometric variations have also been observed in 
NMNLs. For example, RW Tri was observed by Honeycutt et al. (1994b) to show
sinusoidal-like brightness variations of 0.5 magnitude with a period of 25
days. RW Tri and UX UMa also show aperiodic variations in their orbital period
(Still et al. 1995a; Rubenstein et al. 1991). There have been suggestions that
these phenomena might be the result of magnetic cycles within the secondary
star -- see Applegate (1992) and references therein. 

Moving to variations on orbital timescales, the light curves of
high-inclination NMNLs exhibit smooth, round-bottomed eclipses, often with a
pronounced egress shoulder and orbital hump due to the bright spot (see
figure~5). Similar orbital light curves are exhibited by dwarf novae in
outburst. The eclipse light curves of NMNLs can be variable from cycle to
cycle, both in depth and in shape (eg. PX And; Thorstensen et al. 1991).
Eclipse maps show that the discs are generally symmetric, much brighter in
their centres than their edges, and have bright spots whose strengths vary
considerably from object to object (Rutten et al. 1992). 

\begin{figure}[t]
\begin{picture}(100,237)(50,30)
\put(0,0){\includegraphics{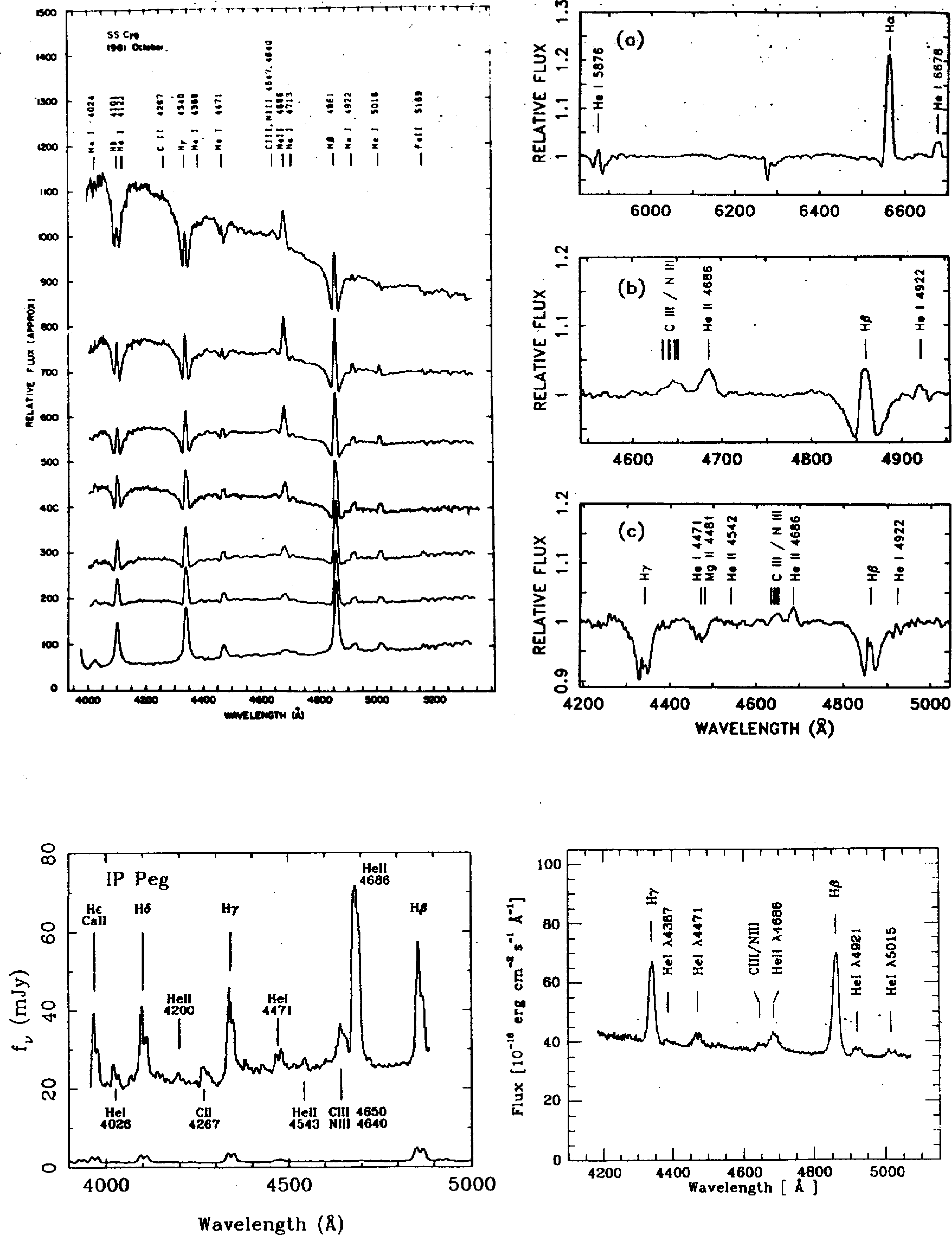}}
\noindent
\end{picture}
\caption{A comparison of the spectra of dwarf novae and NMNLs. Clockwise from
top left: the low-inclination dwarf nova SS Cyg in outburst and quiescence
(Hessman et al. 1984); the low-inclination UX UMa-star IX Vel (Beuermann \&\
Thomas 1990); the high-inclination SW Sex-star WX Ari (Beuermann et al. 1992);
the high-inclination dwarf nova IP Peg in outburst and quiescence (Marsh \&\
Horne 1990).} 
\end{figure}
On timescales of minutes, all NMNLs show stochastic flickering with amplitudes
of up to several tenths of a magnitude. By analysing the behaviour of the
flickering in RW Tri during eclipse, Horne \&\ Stiening (1985) showed that the
entire disc participates in the flickering; it is not known whether this
flickering is the result of irradiation from near the white dwarf, or energy
released at localized sites in the disc. On even shorter timescales,
quasi-periodic oscillations (QPOs) with periods of tens of seconds are
sometimes seen in the light curves of NMNLs (and dwarf novae during outburst),
eg. V3885 Sgr (Warner 1973) and UX UMa (Nather \&\ Robinson 1974). Petterson
(1980) successfully modelled the systematic changes observed in the QPO phases
during eclipse with an accretion disc reflecting radiation from a rotating
source on or near the surface of the white dwarf. Although various models for
the source of radiation have been invoked (eg. bright spots on the white dwarf;
non-radial pulsations; bright spots at the inner edge of the accretion disc),
their true nature remains uncertain. 

\begin{figure}[t]
\begin{picture}(100,124)(50,100)
\put(0,0){\includegraphics{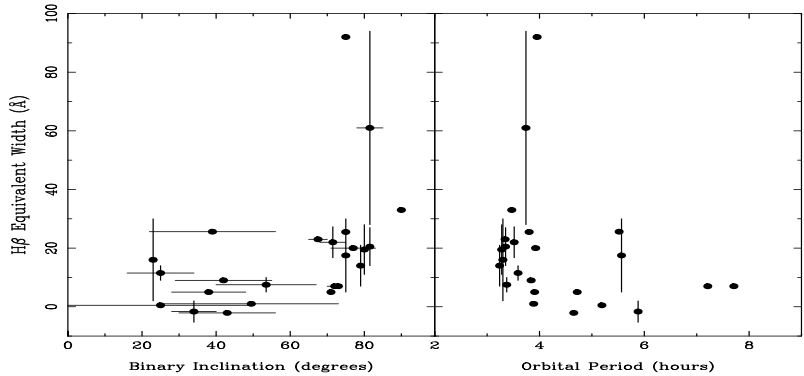}}
\noindent
\end{picture}
\caption{Binary inclination and orbital period versus the equivalent width of
H$\beta$ in a sample of 24 UX UMa, VY Scl and SW Sex-type nova-like variables.}
\end{figure}
NMNLs exhibit the same variety in their optical spectra as do dwarf novae in 
outburst and quiescence, ranging from pure emission-line to almost pure 
absorption-line spectra. In figure~3, representative spectra of high and 
low-inclination dwarf novae in outburst and quiescence are displayed alongside
spectra of high and low-inclination NMNLs. The similarities are striking. 
Low-inclination dwarf novae in outburst show strong absorption lines, as do 
low-inclination NMNLs. High-inclination dwarf novae in outburst show strong 
emission lines, as do high-inclination NMNLs. This strong correlation between 
orbital inclination and the strength of the emission lines has been 
demonstrated for classical novae by Warner (1987) and has been noted for dwarf
novae in outburst by Robinson et al. (1993). Figure~4, which plots the 
equivalent width of H$\beta$ versus the orbital inclination, shows that the 
same correlation exists in NMNLs; low-inclination NMNLs generally show 
absorption lines (often with strong emission cores, which is why the 
equivalent widths are rarely negative), whereas high-inclination NMNLs show 
emission lines. The reason that this correlation exists is due to dilution: at
high inclinations the flux in the optically-thick continuum of the accretion 
disc is reduced by projection and limb darkening, so there is less dilution of
the emission-line flux. Figure~4 also shows that the equivalent width of 
H$\beta$ increases with decreasing orbital period in NMNLs. Hessman (1985) and
Echevarria (1988) found a similar correlation exists for all CVs. This effect 
can be understood in terms of the correlation between line strength and 
absolute magnitude found by Patterson (1984): as the accretion disc becomes 
brighter the emission lines are diluted by the optically thick continuum flux 
from the disc and hence their equivalent widths decrease. Since CVs with 
longer orbital periods tend to have higher rates of mass transfer (see 
figure~2), their emission-line strengths decrease. 

\begin{figure}[t]
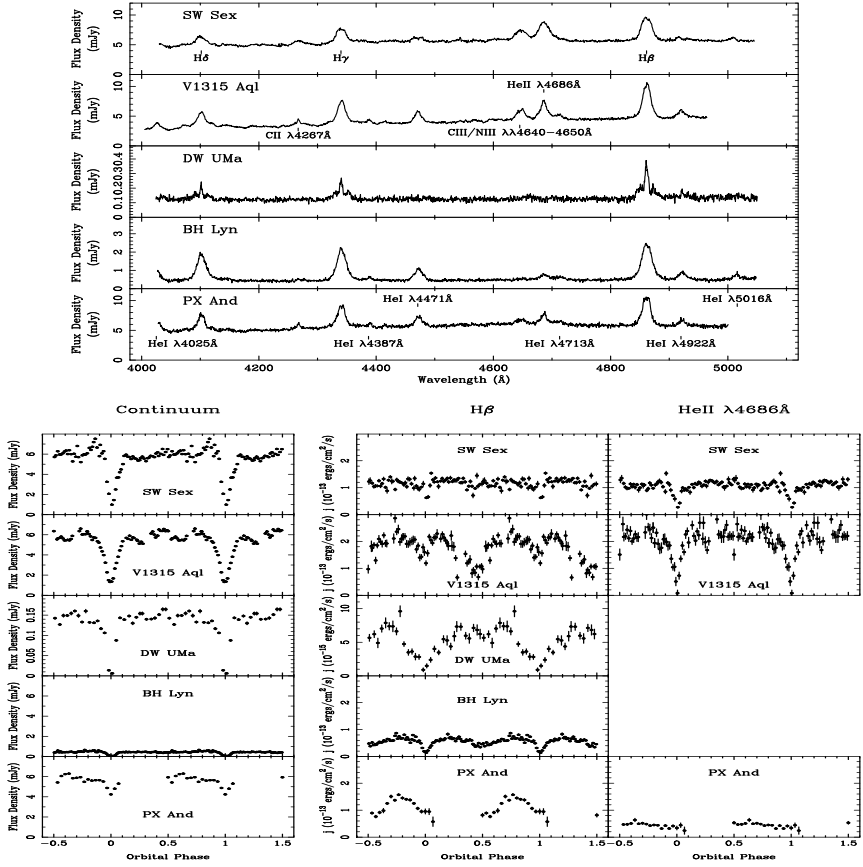

\begin{picture}(100,314)(50,40)
\put(0,0){\includegraphics{swsex_spectra.ps}}
\put(0,0){\includegraphics{swsex_lc.ps}}
\noindent
\end{picture}
\caption{Spectra and light curves of 5 SW Sex stars. From top to bottom: SW Sex
(Dhillon 1990), V1315 Aql (Dhillon et al. 1991), DW UMa (low state; Dhillon et
al. 1994), BH Lyn (low state; Dhillon et al. 1992) and PX And (Still et al.
1995b).} 
\end{figure}
The remainder of this section will review the spectroscopic properties specific
to the UX UMa, SW Sex and VY Scl sub-classes. Perhaps the best understood class
of NMNLs are the UX UMa stars. That this is the case has been verified in some
detail by the spectral eclipse-mapping results of Rutten et al. (1994). They
reconstructed the spectra of different parts of the accretion disc in UX UMa
and found an inner disc and bright spot with a blue continuum and Balmer
absorption lines and an outer disc with a red continuum and Balmer emission
lines. This is qualitatively what is expected from accretion discs which are
hot and optically thick in their centres, and cool and optically thin in their
outer regions (Williams 1980). In addition, theoretical spectra now appear to
be giving encouraging fits to the observed continua of UX UMa stars (Shaviv \&\
Wehrse 1993). 

The same agreement with theory cannot be said to exist for SW Sex stars. This
group of stars are all high-inclination NMNLs\footnote{\tiny Even WX Ari,
previously thought to be low inclination, shows grazing eclipses (Hellier et
al. 1995).} with periods in the range 3--4 hr. The spectra of 5 SW Sex stars
and their continuum and emission-line light curves are displayed in figure~5. 
Note that the spectra and light curves of DW UMa and BH Lyn were obtained 
during low states (see below). All SW Sex stars exhibit strong, single-peaked 
Balmer, HeI and HeII emission lines which, with the exception of HeII, remain 
largely unobscured during primary eclipse. This is in stark contrast to 
standard accretion disc theory which predicts that emission lines from 
high-inclination discs should appear double-peaked and be eclipsed once every 
orbital period. In addition, the emission lines in SW Sex stars show strong 
absorption features around phase 0.5 and their radial velocity curves exhibit 
significant phase shifts relative to photometric conjunction. As shall be 
discussed in section~4, these observational constraints have yet to be fully
satisfied by a single theoretical model. 

The VY Scl stars in their normal state are indistinguishable from UX UMa or SW
Sex stars. In their low states, however, which are due to a decrease in the
mass accretion rate, VY Scl stars completely change in appearance. An example
of typical VY Scl-star behaviour has been presented by Dhillon et al. (1994),
who observed DW UMa in a $\sim 3$ magnitude low state; the resulting spectrum
and light curves are shown in figure~5. It can be seen that, in contrast to the
normal state, there are no high-excitation emission lines present. The Balmer
lines are dominated by strong, narrow emission spikes superposed upon faint
wings. Dhillon et al. (1994) showed that the emission spikes originate on the
secondary star and the line wings show evidence for an accretion disc origin.
By measuring the continuum eclipse width, they also found that the accretion
disc was smaller than during the normal state. Using this same data in
conjunction with Roche tomography, Rutten \&\ Dhillon (1994) were able to map
the Balmer-line intensity distribution on the surface of the secondary in DW
UMa and showed strongly enhanced emissivity on the inner face of the secondary
star. A similar result was also obtained by Still et al. (1995a) in their 
study of RW Tri during a $\sim 3.5$ magnitude {\em high} state, suggesting that
irradiation by the accretion disc plays an important role in NMNLs. 

\section{Theoretical Models}

The observed similarities between NMNLs and dwarf novae in outburst suggest
that NMNLs might be successfully modelled by canonical CVs with steady-state 
accretion discs in which the mass transfer rate is sufficiently high to prevent
disc instability-type outbursts. But can this model explain all of the 
observed NMNL phenomena, and in particular, the low states, the orbital period
distribution, the departures from steady-state accretion in the inner disc 
regions and the SW Sex phenomenon? In this final section we will show that a 
number of modifications to the canonical model are required in order to 
approach an understanding of the NMNLs, a conclusion which has also been 
reached by la Dous (1993b). 

Turning first to the SW Sex phenomenon, it is clear that a successful model 
will have to include some component of line emission from above the orbital 
plane in order to explain the lack of eclipse in the emission lines. 
For this reason, the Stark broadening model of Lin et al. (1988)
and the bright-spot overflow model of Hellier \&\ Robinson (1994) are unlikely
to be correct. However, the latter model is very successful in explaining many 
of the other properties of SW Sex stars, such as the phase 0.5 absorption.
This suggests that a model with a bright-spot overflow operating in 
conjunction with some form of accretion disc wind (Honeycutt et al. 1986; 
Hoare 1994; Dhillon \&\ Rutten 1995), magnetic accretion column (Williams 
1989), or magnetically driven outflow (Tout et al. 1993; Wynn et al. 1995) 
might be able to explain all of the observed phenomena. The requirement that 
some mechanism drives material out of the orbital plane is also consistent 
with the observed departures from steady-state accretion in the inner discs
of these systems and provides an explanation for the redder UV spectra of 
NMNLs compared to dwarf novae in outburst (Tout et al. 1993). 

The first attempt to explain the observation of low states in VY Scl stars
was made by Robinson et al. (1981). They noted the period grouping of VY Scl
stars and speculated that the low states are a consequence of their imminent 
entry into the period gap. In this model, the magnetic braking which drives
mass transfer ceases as a result of the secondary star becoming fully
convective at periods immediately above the gap (Spruit \&\ Ritter 1983). 
However, Livio \&\ Pringle (1994) showed that this model cannot account for
the low states in VY Scl stars due to the disparate timescales -- it takes
VY Scl stars of order 10--100 days to enter a low state whereas it would take 
10000 years or more to respond to a sudden cessation in the mechanism driving 
the mass transfer. They proposed an alternative model where star-spots 
covering the L$_1$ point block the mass transfer and hence cause low states. 
It may be possible to test this model using Roche tomography (Rutten \&\ 
Dhillon 1994).

Using the eruptive characteristics of CVs to infer relative mass transfer
rates, Shafter (1992) concluded that magnetic braking models have severe
difficulties accounting for the orbital period distribution of CVs, in
particular the observed dearth of dwarf novae with orbital periods immediately
above the period gap and the dominance of NMNLs in this same period range.
Intriguingly, the star-spot model of Livio \&\ Pringle (1994) can also be used
to explain this orbital period distribution. The dwarf novae, with lower mass 
transfer rates, are more easily interrupted by the star-spot mechanism because
the density and pressure are correspondingly lower at the L$_1$ point. As the 
periods decrease, the magnetic fields and their covering factors might be 
expected to increase, so that if mass transfer occurs it can only do so at a 
relatively high rate. In their picture, therefore, the low mean mass transfer
rate systems (dwarf novae) become detectable as high mass transfer rate 
systems (NMNLs) as they approach the period gap. 

But why do NMNLs have higher mean mass transfer rates than dwarf novae? This 
is ultimately the most important question of all, the answer to which remains
uncertain. According to the hibernation model for the cyclical evolution of
CVs (see Duerbeck 1993 and references therein), it is due to the fact that 
NMNLs are post-novae whose outbursts we have missed. Only time will tell if 
this theory is correct.

\end{document}